\documentclass[pdflatex,sn-nature]{sn-jnl}

\usepackage{graphicx}%
\usepackage{multirow}%
\usepackage{amsmath,amssymb,amsfonts}%
\usepackage{amsthm}%
\usepackage{mathrsfs}%
\usepackage{xcolor}%
\usepackage{textcomp}%
\usepackage{manyfoot}%
\usepackage{booktabs}%
\usepackage{algorithm}%
\usepackage{algorithmicx}%
\usepackage{algpseudocode}%
\usepackage{listings}%
\usepackage{chngcntr}
\usepackage{makecell}
\usepackage{indentfirst}

\theoremstyle{thmstyleone}%
\theoremstyle{thmstyletwo}%

\theoremstyle{thmstylethree}%

\newcommand\subsetsim{\mathrel{\substack{
    \textstyle\subset\\[-0.2ex]\textstyle\sim}}}

\begin{document}

\title[Optimizing Research Portfolio]{Optimizing Research Portfolio For Semantic Impact}


\author*[]{\fnm{Alexander} V. \sur{Belikov}}\email{alexander@growgraph.dev}

\affil*[]{\orgname{Growgraph}, \orgaddress{
    \city{Paris}, \postcode{75008}, \state{Île-de-France}, \country{France}}}



\abstract{
    Citation metrics are widely used to assess academic impact but suffer from social biases, including institutional prestige and journal visibility.
    Here we introduce rXiv Semantic Impact (XSI), a novel framework that predicts research impact by analyzing how scientific semantic graphs evolve in underlying fabric of science.
    Rather than counting citations, XSI tracks the evolution of research concepts in the academic knowledge graph (KG).
    Starting with a construction of a comprehensive KG from 324K biomedical publications (2003-2025), we demonstrate that XSI can predict a paper's future semantic impact (SI) with remarkable accuracy ($R^2$ = 0.69) three years in advance.
    We leverage these predictions to develop an optimization framework for research portfolio selection that systematically outperforms random allocation.
    We propose SI as a complementary metric to citations and present XSI as a tool to guide funding and publishing decisions, enhancing research impact while mitigating risk.
}

\keywords{Knowledge Graphs, Impact Metric, Portfolio Optimization, Science of Science}

\pacs[JEL Classification]{D81, H51}

\pacs[MSC Classification]{91G10, 91D30}

\maketitle

\section{Introduction}\label{sec:intro}

Every year enormous research budgets are allocated across competing research proposals.
The decisions of funding some projects, at the expense of others, shape the trajectory of scientific progress and, ultimately, the ability of humankind to address challenges from disease to climate change.
Yet the tools we use to evaluate research impact remain remarkably crude - primarily relying on citation counts that often reflect social dynamics more than scientific merit (\cite{McNutt2014-hh, Burton2024-ns}).

Consider a breakthrough paper proposing a novel genetic treatment.
Under the current system, it may be misjudged by the community due to the publishing journal's prestige or the authors' institutional affiliations, rather than the scientific content itself.
Even after publication, it could take years to accumulate enough citations to be recognized as influential, particularly if it was published in a less prominent journal or from researchers at less known institutions.

This dependence on citation-based metrics creates systematic inefficiencies allocation of scientific resources.
High-quality research from less known centers may be overlooked, while work from established groups might receive disproportionate attention.

%

Venture capital, responsible for stimulating the implementation and adoption of scientific innovation, also suffers from an ensemble of biases~\cite{Sachs2024-tb}, but lacks the experience to evaluate deep-tech opportunities.
In both academic and business settings, machine-driven, semantically enabled tools trained in academic knowledge networks could greatly boost the success of discovery and adoption of scientific innovations.

While metrics of scientific quality beyond simple citation counts were proposed, for example defining high impact as belonging to the top 5\% centrality nodes of the citation network, see~\cite{Weis2021-yo}, recent advancements in the large language models (LLM) open new avenues for analyzing scientific literature.
By leveraging LLMs, it is now possible to trace the emergence of novel ideas and model their future evolution at scale by constructing semantic networks of concepts, known as knowledge graphs (KGs).
These KGs represent ideas and their interconnections providing a framework to assess the influence of specific subgraphs on the future landscape of scientific knowledge.

In this context, we propose Semantic Impact (SI), a novel metric designed to capture how research findings affect the future knowledge network.
This approach aims to provide a more direct measure of a publication's impact, addressing the limitations associated with citation-based evaluations.


We present the following key contributions: (i) we construct the most detailed knowledge graph to date from 324K research publications from arXiv, bioRxiv, and medRxiv; (ii) we introduce Semantic Impact as a metric that captures the structural influence of a publication within the evolving knowledge network and demonstrate its statistically significant correlation with the citation counts; (iii) we analyze the effects of COVID-19 on the global KG dynamics; and (iv) we develop a predictive framework for SI and propose a portfolio optimization approach that selects research publications to maximizes their future semantic impact.
We conclude by a discussion of potential applications of this framework, its implications for research assessment, and directions for further improvements.

\begin{figure}[h]
    \centering
    \includegraphics[width=1.\textwidth]{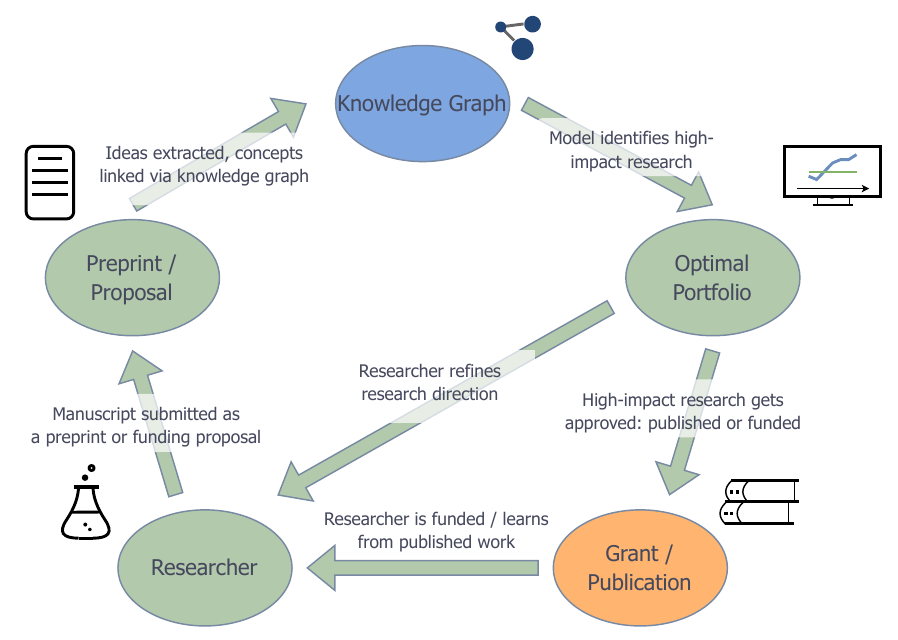}
    \caption{
        \textbf{The positive feedback loop in research optimization}:
        Preprints and research proposals enter the Semantic Impact System (XSI), where they are transformed into semantic knowledge graphs.
        Predictive models assess their future semantic impact and associated risks.
        This information guides researchers in shaping their work and aids publishers and funding agencies in making data-driven decisions on publications and support.
        These decisions, in turn, accelerate the growth of the global academic knowledge graph.
    }\label{fig:feedback.loop}
\end{figure}

We envision a positive feedback system in which research proposals and published work continuously enrich a global knowledge graph.
This structured representation enables a more optimal and unbiased selection process for funding and publishing, prioritizing projects with the highest predicted future impact (see Fig.~\ref{fig:feedback.loop}). Even a modest improvement in each iteration of this cycle could drive exponential growth in academic knowledge.

\section{Results}

\subsection{Knowledge Graph Preparation}
\label{subsec:kg}

We focus on open access publications related to biology and medicine: preprints posted on biorXiv\footnote{https://biorxiv.org}, medrXiv\footnote{https://medrxiv.org} and arXiv\footnote{https://arxiv.org} (only submissions categorized as quantitative biology \textit{q-bio}), totaling 324K as of December 24, 2024.

The abstracts are parsed using TriEL, an NLP engine designed for relation extraction and entity linking.
Relation extraction is a standard NLP task that identifies structured relationships in text by extracting triples of the form (subject, relation, object).
For example, a fact such as \textit{"The Eiffel Tower is in Paris"} would be represented as (the Eiffel tower, is located in, Paris).
Entity linking assigns unique identifiers to words or phrases, mapping them to entities in a knowledge base.
TriEL, based on SpaCy~\cite{spacy}, processes text through the following steps: (i) identifying \textit{candidate} subject, object and relation (referred to as \textit{mentions}), (ii) proposing triples based on graph distances between candidates, (iii) running co-reference resolution, linking equivalent mentions, and (iv) executing entity linking using self-hosted services BERN2~\cite{sung2022bern2} and Entity Fishing~\cite{entity-fishing}.

If a mention cannot be mapped to an existing entity in the knowledge graph, we generate a compact identifier using the following steps: (i) lemmatizing the constituent words, (ii) excluding high-frequency words, if multiple words remain, (iii) merging the remaining words, deriving a hash code and assigning entity type $T_0$.

Following these steps we generate a directed knowledge graph of entities $G_{\pi}$ from a research paper abstract $\pi$, where entities and relationships between them are explicitly captured.
This method provides a more precise representation compared to previous works, where relationships were inferred from co-occurrence statistics~\cite{venugopal2022, Krenn2023-br}.

To facilitate the analysis, we aggregate individual publication graphs into a global knowledge graph: $G = \bigcup\limits_{\pi} G_{\pi}$ and a KG sequence $\{G_r(t)\}$ by removing non-informative vertices (e.g.~type $T_0$) and retaining only edges from publications before time $t$ (see stages B and C, Fig.~\ref{fig:flow}).
Please refer to Appendix~\ref{sec:appendix:a} for further details.

We note a dip in the number of publications following 2020 (Fig.~\ref{fig:kg.stats}, top row), which we attribute to the consequences of COVID-19.

\subsection{Semantic Impact}\label{subsec:impact}

We introduce a novel knowledge graph metric to quantify the importance of a subgraph $\pi$, issued at time $t$, with respect to a sequence of evolving knowledge graphs $G_r(t)$.
We call this metric Semantic Impact (SI), denoted as $J_\pi$.
SI captures how the popularity of a subgraph changes over time, measuring its importance from its initial graph $G_r(t)$ to a future $G_r(t + \Delta)$.
It is defined as the average ratio of edges weights in the future graph $G_r(t + \Delta)$ normalized by those at the time of publication $G_r(t)$, where $\Delta$ represents for prediction horizon.
To ensure that the average $J_\pi$ evolves approximately linearly over time, we apply a $\ln (1 + \star)$ transformation in the final step.

We evaluate $G_r(t)$ graphs for end of month dates, which enables a practical way to measure intervals in integers of months.

SI satisfies the following two edges cases: (A) a publication that is popular at time $t$, while not popular at $t + \Delta$ will have low impact, (B) a subgraph that is not popular at time $t$ and popular at $t + \Delta$, on the contrary, will have high impact.

Having defined SI and trained a model for its prediction we also define its relative prediction error $\delta_\pi = \left | \frac{J_\pi - \hat{J_\pi}}{J_\pi}\right|$.
Between two publications with equal predicted semantic impact we should prefer the one with a smaller SI prediction error.

\begin{figure}[h]
    \centering
    \includegraphics[width=1.\textwidth]{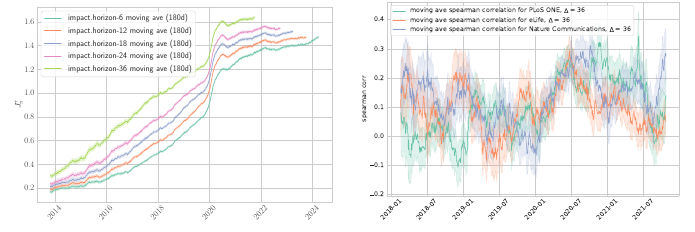}
    \caption{\textbf{Left}: moving average of SI $J_\pi$ with a window of 180 days. \textbf{Right}: Spearman correlation of SI $J_\pi$ and citation counts from Openalex over a 90 day window for the total citation count as of Openalex data fetch.}\label{fig:scale_citations}
\end{figure}

We observe that semantic impact is non-stationary and consider the prediction of SI with respect to its moving average with a window of 180 days (Fig.~\ref{fig:scale_citations}, left).

To corroborate the definition of semantic impact, we compute the Spearman correlation of $J_\pi$ with total citation counts for papers published in peer-reviewed journals (data provided by Openalex) in a 90-day window (see Fig.~\ref{fig:scale_citations}, right).
We observe that the SI/citation count correlation, grouped by journals to exclude journal bias, is positive and vary in the range [0.0,0.3] with an average value of $0.11$ for PloS One, $0.13$ Nature communications, and $0.09$ for eLife, the journals with the greatest number of publications with citation data in our sample.

In order to predict semantic impact $J_\pi$ of a publication $\pi$ for example 3 years in the future, we consider only semantic features, available at the time of publication.
We derive the following types of features for subgraph $\pi$ with respect to the graph $G_r(t)$: originality $\omega_\pi$, defined as a normalized average weight of edges of the subgraph (see Eq.\ref{eq:originality}); mean and dispersion of in- and out- vertex degrees; community-related and graph diffusion metrics.
Please refer to Appendix~\ref{sec:appendix:b} for more details.

In the sampling procedure we start with a sequence of end of month dates $\{t_i\}$ and pick publications from the period $[t_i - T_{train}, t_i]$, on which we train the model, and validate the model on publications from $[t_i + \Delta , t_i + \Delta + T_{valid}]$.
Therefore, for an evaluation date of 31 December 2024, we can calculate the 36-month SI for publication no earlier than December 2021 and so on.

\begin{figure}[h]
    \centering
    \includegraphics[width=0.9\textwidth]{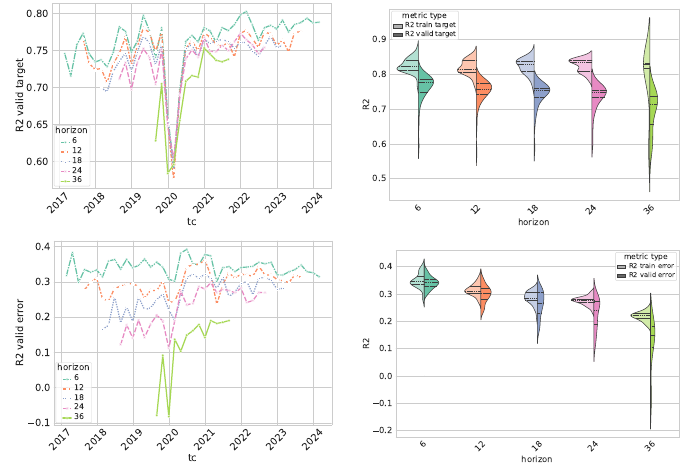}
    \caption{Left: line plots, representing coefficient of determination $R^2$ for the validation sample as a function of time. Right: violin plots of the distributions of coefficient of determination $R^2$ for the training and validation sample (left and right halves, lighter and darker hues of the same color correspondingly). The target is transformed, both the target and the features are scaled. The training period is 36 months. The cases represent the target and different prediction horizons $\Delta = 6, 12, 18, 24$ and 36 months.}\label{fig:model.perf.horizon}
\end{figure}

Following our preliminary analysis, we chose to model normalized semantic and use normalized features for model evaluation (Fig.~\ref{fig:model.perf.sf}).
We found that the longer training period, the better were the out-of-sample coefficient of determination (Fig.~\ref{fig:model.perf.train.period}).
As the final step we identify model parameters that delivered the best out-of-sample performance per prediction horizon (Fig.~\ref{fig:model.perf.horizon}) for $\Delta = 6,12,18, 24, 36$ months, where the predictive power of the target variable (as well as the prediction of its error) degrades for greater horizons.

We note the presence of a performance trough at the beginning of 2020, attributed to COVID-19 (see Appendix~\ref{sec:appendix:c}).

\subsection{Optimization Framework}\label{subsec:opt}

In much the same way that an investor selects $n$ assets to construct a portfolio that maximizes performance over a given time horizon $\Delta$, funding agencies and journals must choose $n$ of the most promising projects or publications.
Likewise, research labs and individual researchers allocate their resources and time with the expectation of returns, such as citations, grants, or enhanced standing within the academic community.
In both cases, decision-makers distribute limited resources, whether time, funding, or journal space, to maximize measurable outcomes, such as financial returns in investment or academic impact in research.

This process bears an analogy to the Markowitz portfolio optimization problem~\cite{Markowitz1952-vh}, where at time $t$ we seek to select publications that will maximize the semantic impact at a future time $t + \Delta$, while minimizing the variance of this prediction.
However, several key differences set our problem apart from standard portfolio optimization: (i) at each time step,  we make a selection among newly published works, requiring a predictive model for $P(J_\pi | X)$, rather than relying on historical stock performance; (ii) while inter-publication dependencies exist (e.g., citations shared between similar papers), we initially assume publication performances to be independent, (iii) Unlike stocks, which can be acquired in fractional or variable amounts, each publication is either selected or not, (iv) the absence of the risk-free asset, (v) the inability to take short positions on research publications.

With these distinctions in mind we formulate an optimization problem that aims to maximize the expected impact $J_\pi$ while minimizing its predicted uncertainty.
The problem is a variant of the 0--1 knapsack problem and is solved using Integer Linear Programming in the following optimization framework:
\begin{align}
    \max \sum\limits_{\pi\in \Pi} x_\pi(J_\pi - \beta\delta_\pi)  \nonumber \\
    \mbox{such that} \frac{1}{|\Pi|} \sum\limits_{\pi \in \Pi} x_\pi \le \alpha\\
    \mbox{where}\: x_\pi \in \{0, 1\}, \forall \pi \in \Pi, \nonumber
\end{align}

where $\Pi$ is the set of publications available for the selected time period, $J_\pi$ represents the predicted impact of publication $\pi$, $\delta_\pi$ denotes the associated uncertainty or risk, $x_\pi$ is a binary decision variable indicating whether publication $\pi$ is selected,  $\alpha$ is the maximum fraction of selected publications, and $\beta$ is a model parameter, allowing to prioritize lower variance for its higher values.

We use OR-Tools~\cite{cpsatlp} to solve the plan above.

\begin{figure}[h]
    \centering
    \includegraphics[width=0.9\textwidth]{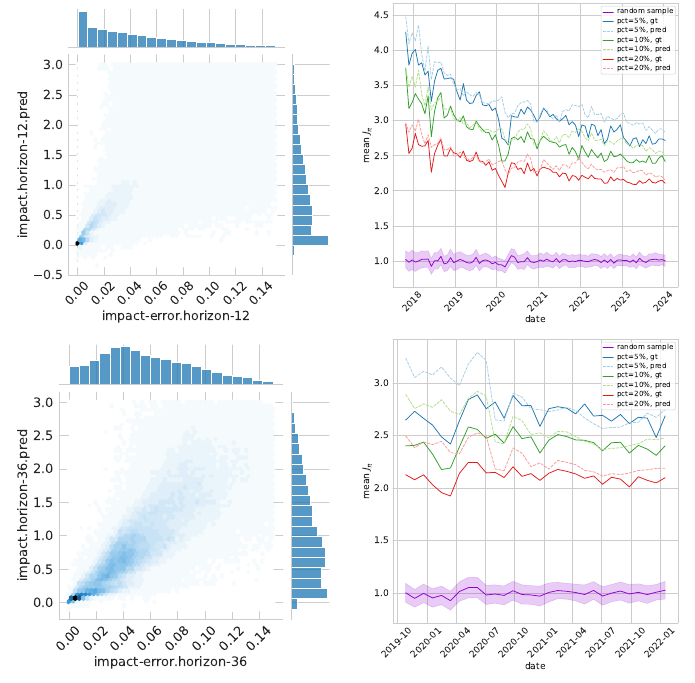}
    \caption{Left: SI $J_\pi$ - SI prediction error $\delta_\pi$ plot for prediction horizons $\Delta$ = 12 months (top) and 36 months (bottom). Right: Optimal portfolio performance for prediction horizons $\Delta$ = 12 months (top) and 36 months (bottom). The top 3 lines on both plots correspond to the selection of $5, 10, 20\%$ of available publications for the period, the bottom line (blue) with the band denoting the standard deviation error, corresponds to a random selection of publication available for each }\label{fig:opt}
\end{figure}

The models developed in Appendix~\ref{sec:appendix:a} enable portfolio selection using a temporal out-of-sample validation (back-testing) (see Fig.~\ref{fig:opt}).
We observe that, as expected, the realized portfolio consistently underperforms the predicted portfolio.
Since SI is normalized, the performance of a random selection of publications remains approximately unity.
Unlike the mean-variance framework, where the assumption of normal risk distributions produces a sharply defined efficient frontier, in our model the efficient frontier is smooth.

We note that compared to the traditional mean-variance optimization framework our models are not normal and thus the efficient frontier is not strict.

\section{Discussion}\label{sec:discussion}

To demonstrate the utility of semantic metrics, we developed a procedure for constructing the most comprehensive biomedical knowledge graph of preprint servers to date.

Although previous studies have relied on high-quality annotated resources for precision medicine~\cite{Chandak2023-dd} or analyzed co-occurrences of key concepts~\cite{Venugopal2024-el, gu2025}, our approach focuses on extracting candidate triples and linking them to entities.
This enables a more nuanced analysis of both the semantic and directional aspects of relationships.

To demonstrate the utility of semantic metrics, we developed a procedure for constructing the most comprehensive biomedical knowledge graph of preprint servers to date.
While previous studies have either relied on high-quality annotated resources for precision medicine [12] or analyzed co-occurrences of key concepts [13, 14], our approach focuses on extracting candidate triples and linking them to entities.
This enables a more nuanced analysis of both the semantic and directional aspects of relationships.

The semantic impact metric, while more objective than citation-based measures in certain respects, has inherent limitations.
It primarily reflects the declarative quality of research projects but does not account for how well the study was executed.
For example, SI might not distinguish between a publication that only presents a hypothesis and one that provides a complete proof.
Similarly, SI favors first-mover advantage: an earlier publication stating a claim will receive a higher score than a later publication presenting the actual evidence.
Our current implementation also has technical limitations.
Firstly, while repositories such as arXiv, bioRxiv, and medRxiv are extensive, they do not offer comprehensive coverage of biomedical sciences.
Secondly, the precision of our entity linking tools is limited.
For example, semaglutide (Ozempic) may be incorrectly mapped to semaphorin 5B.
Beyond these imperfections, entity linkers are trained on data available at the time of training and may not perform well on newly emerging entities.

Since the biases of citation-based metrics, such as institutional prestige, publication venue, and researchers’ visibility, are orthogonal to those of SI, the latter can serve as a complementary measure of publication quality.

We observe that the accuracy of the impact prediction decreases as the prediction horizon increases, while the variance of the predictions increases.
A similar trend is evident in modeling prediction error.
Additionally, COVID-19 disrupts the form of SI distribution since external factors made COVID-related statements more prominent, as shown in Table~\ref{table:covid} and Fig.\ref{fig:covid}, (see Appendix ~\ref{sec:appendix:c} for an extended discussion).
This disruption increasing the complexity of the prediction task (see the performance drop in Fig.\ref{fig:model.perf.horizon}).

Future improvements include expanding the dataset to construct a more comprehensive knowledge graph, incorporating sources such as PubMed.
Enhancing entity linking precision and processing full-text publications instead of abstracts would also improve the model's performance.
To better capture the nuances of semantic impact, it would be beneficial to classify knowledge graph elements according to their rhetorical function, such as background, objectives, data, experimental design, methods, results, and conclusions, following~\cite{prabhakaran.predicting.2016}.

From a modeling perspective, the methods presented in this work are both robust and interpretable, owing to their well-defined feature sets.
However, deep learning approaches, particularly Graph Neural Networks (GNNs), could further enhance predictive performance by capturing richer graph representations.

Another promising direction is to integrate scientometric metadata, such as authorship and institutional affiliations, into the knowledge graph.
This would allow us to analyze the sociological factors that influence the semantic evolution of science.
It could also enable us to assess the correctness of scientific claims, measuring the likelihood that a publication’s statements hold up over time, in line with~\cite{Belikov2022-gj}.

Finally, XSI may also be applied for effective hypothesis generation in the spirit of ~\cite{Sourati2023-zf}.
Rather than predicting new links between existing concepts, it may be used to generate new complex proposal, for example the combination of disease, drug and target~\cite{Jia2024-bd}, crucial for treatment of neglected diseases.

The XSI framework is generalizable and centers on a specific formulation of semantic impact, capturing semantic popularity and extending it with a portfolio optimization model that maximizes SI while minimizing prediction risk.
Different stakeholders - such as research institutions, universities, publishers, and funding agencies - may have distinct definitions of what constitutes a successful scientific contribution.
They may wish to refine the semantic impact metric according to their priorities, such as bias mitigation or emphasize either specialization or broad coverage of research topics.

\section{Acknowledgements}
\label{sec:ack}
We thank V. Sitnik, T. Garduño, I. Rushkin and A. Rzhetsky for fruitful discussions and S. Kudinov for help with setting up the web demonstration.

\section{Additional information}
\label{sec:addinf}

\textbf{An interactive demo} of SI/SI error for current (actual) publications posted on arXiv, biorXiv and medrXiv is available at \url{https://growgraph.dev/rxiv-impact}.

\textbf{Extended data} is available for this article at \url{https://zenodo.org/records/14866949}.

\textbf{Codes} is available at \url{https://github.com/growgraph/xsi}.

\bibliography{sn-bibliography}

\newpage

\begin{appendices}

    \section{Methods}\label{sec:appendix:a}

    \begin{figure}[h]
        \centering
        \includegraphics[width=1.\textwidth]{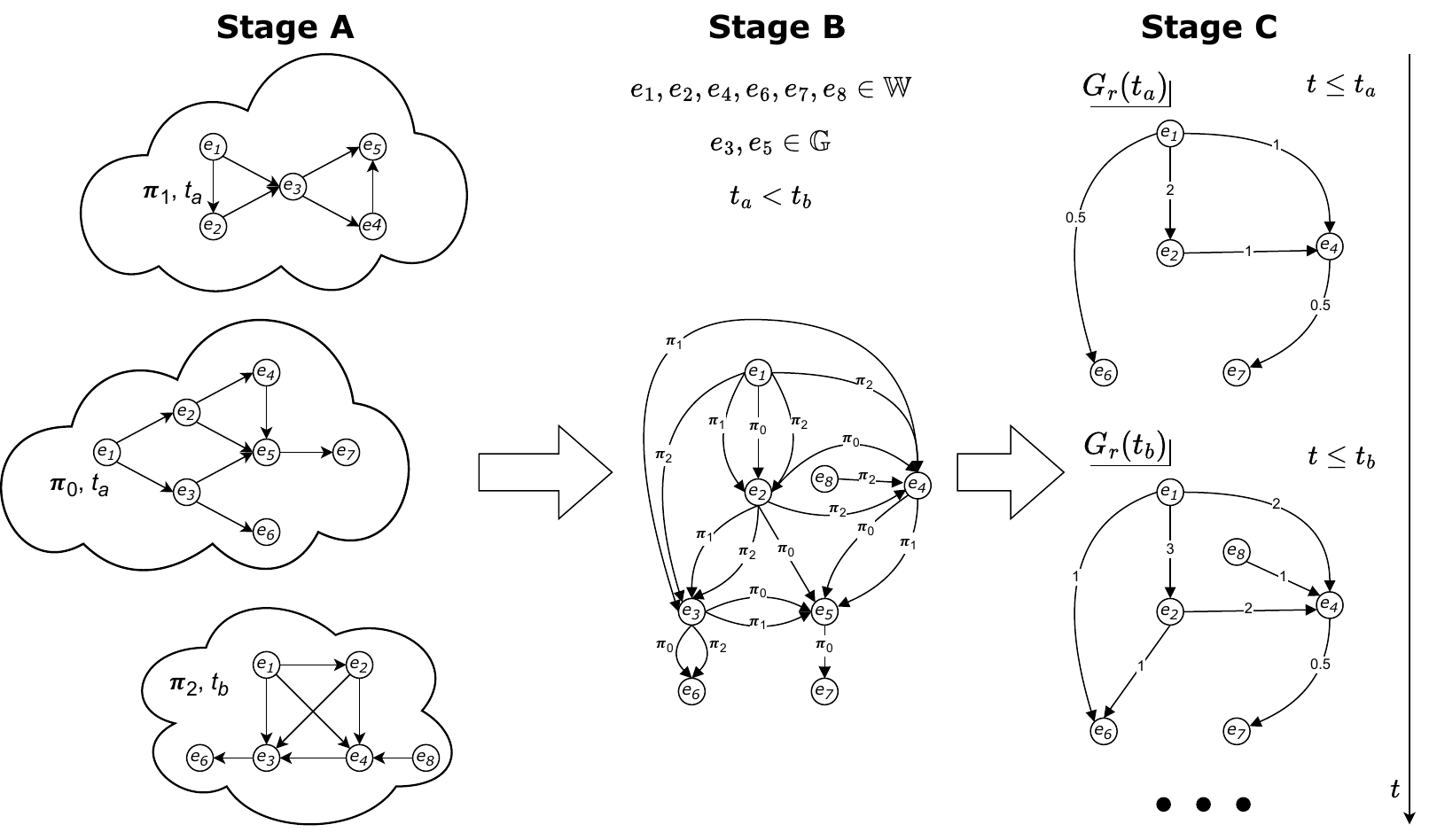}
        \caption{Stage A: each abstract is repsented as a directed graph of entities. Stage B: the union of directed graphs of entites forms a complete domain knowledge graph. Stage C: a series of KGs where each following KG contains the previous directed graphs $G_r(t)$.}\label{fig:flow}
    \end{figure}

    \counterwithin{figure}{section}

    In this section we provide the technical details of KG construction and formal definitions of metrics used for modeling.
    In order to construct graphs $\{G_r(t)\}$, we transform each publication graph in the following way: (i) all edges are initialized with weights of 1, (ii) all vertices of non-informative types (e.g., type $T_0$) are removed, (iii) edges $e_{u\star}$ and $e_{\star u}$, where $\star$ stands for all neighbors of $u$), connected to removed vertices are deleted, and (iv) new edges are introduced between vertices that were connected by a path before non-informative vertex removal with weights defined by $w_{ab}^{-1} = (w_{au}^{-1} + w_{ub}^{-1} )^{-1}$.

    This transformation may introduce new edges (e.g.~edge ($e_2$, $e_6$) after excision of vertex $e_3$).
    For practical reasons we define $G_r(t) = \bigcup\limits_{t - \Delta\le {t_{\pi}\le t}} G_{\{\pi\}}$ with $\Delta = 10\;\mbox{years}$, making the inclusion of $G_r{t_a} \subsetsim G_r{t_b}$ approximate for $t_a < t_b$.
    Thus, $G_r(t)$ encapsulates all domain knowledge available up to time $t$.
    To extract features from $G_r(t)$, we group entities into communities using the label propagation algorithm~\cite{label-propagation}.

\begin{figure}[h]
        \centering
        \hspace*{-1cm}\includegraphics[width=1.2\textwidth]{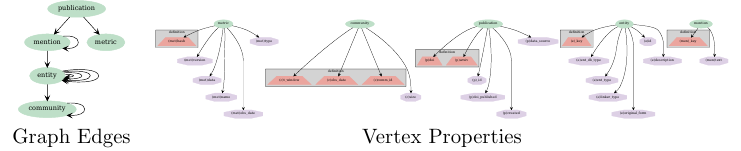}
        \caption{\textbf{Left}: Property graph vertex types and relations between them; \textbf{right}: properties of vertex types.}\label{fig:graph.struct}
    \end{figure}

\begin{figure}[h]
        \centering
        \includegraphics[width=1.\textwidth]{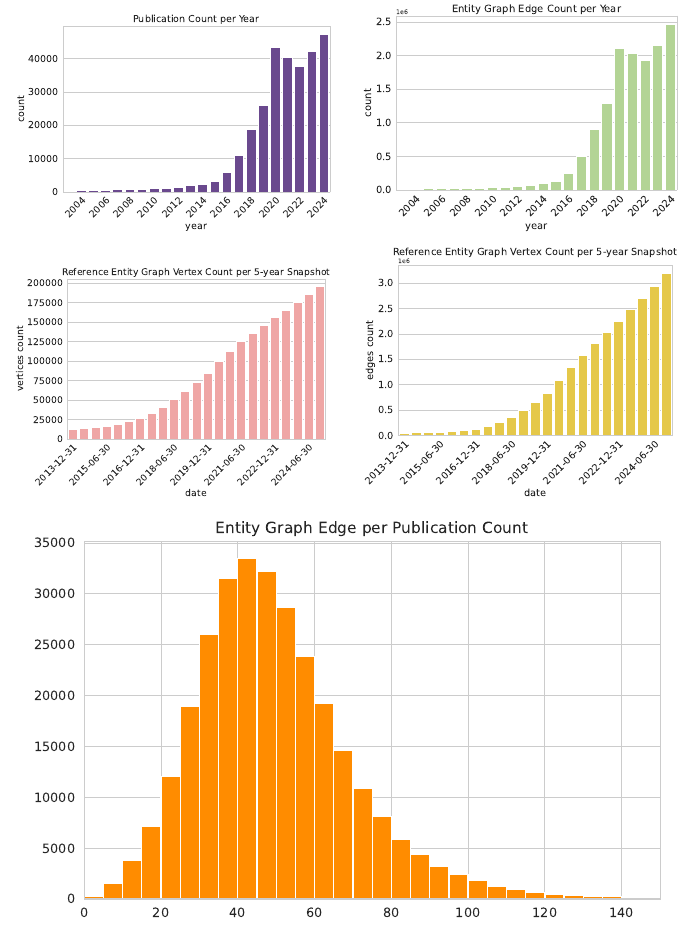}
        \caption{\text{Knoweldge graph statisics}. First row, left: the number of publications present as a function of time, right: the number of edges in the endity graph as a function of year. Second row, left: the vertex number in the entity graph as a function time in the reduced graph $G_r(t)$, right: edge count as a function of time in the reduced graph $G_r(t)$. Bottom row: the distribution of edge count per publication.}\label{fig:kg.stats}
    \end{figure}

    The ensemble of knowledge graphs with vertices representing publications, mentions, entities and communities, itself a knowledge graph is modeled as property graphs and stored in ArangoDB.
    The schema (relations between vertex types and vertex properties) is presented in Fig.\ref{fig:graph.struct}.
    The processing of graphs was facilitated by GraphCast engine, a tool for transforming tabular and json-like data to graphs.
    The properties of the dataset are presented in Fig.\ref{fig:kg.stats}.
    We note that abstracts from 288K publications (or 89\%) rendered non-trivial knowledge graphs, which is due to the fact some publications are not well aligned with respect to the domain of entity linkers.

    We formally define SI as:

    \begin{align}
        \label{eq:def.impact}
        J^0_\pi = \frac{1}{|\pi|} \sum\limits_{e \in \pi} \left ( \frac{w_e (\pi \bigcup G_r(t+\Delta))} {w_e (\pi \bigcup G_r(t))} - 1 \right ),
    \end{align}

    where $e$ is an edge linking entities and $w_e$ is its weight.
    In the text we define SI as $J_\pi = \ln (1 + J^0_\pi)$.

    In case $G_r(t) \subseteq G_r(t + \Delta)$ we have $ w_e (\pi \bigcup G_r(t)) \le w_e (\pi \bigcup G_r(t+\Delta))$ and thus $J_\pi \ge 0$.

    We define \textit{originality} as the inverse of the average edge weight of $\gamma_\pi$ at time $t$:

    \begin{align}
        \omega_{\pi} = \left(\sum\limits_{e \in \pi} {w_e (\pi \bigcup G_r(t))} \right)^{-1}.
        \label{eq:originality}
    \end{align}

    We note that SI moving average for smaller windows is noisy, due to sample size, hence we take 180 days as window size for SI moving average.

\begin{figure}[h]
        \centering
        \includegraphics[width=1.\textwidth]{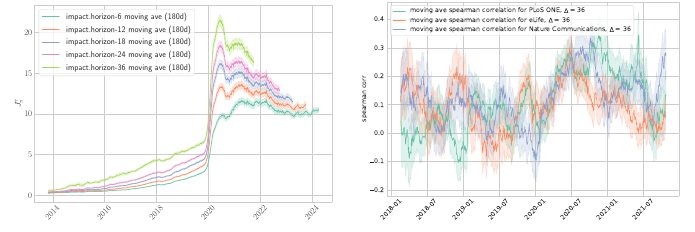}
        \caption{\textbf{Left}: moving average of semantic impact $J^0_\pi$ with a window of 180 days. \textbf{Right}: Spearman correlation of SI $J_\pi$ and citation count from Openalex over a 90 day window: for the total citation count as of Openalex data fetch.}\label{fig:scale_citations.appendix}
    \end{figure}

    By comparing SI moving average Fig.\ref{fig:scale_citations.appendix}, left, and SI moving average with a log transformation applied, Fig.\ref{fig:scale_citations}, left, we conclude than the latter appears to be amenable for modeling using a linear regression, and so we include log-transformed SI in our model evaluation pipeline.
    For temporal validation and prediction outside the historic data interval we extrapolate SI with a linear regression trained on the previous year of data.

    In conclusion, we note that SI distribution manifests an anomalous peak with the left boundary around the end 2019, which we associate with publications related to COVID-19 (see~\ref{sec:appendix:c}).

    Since we expect SI to represent the value of a given publication, we study its correlation with a commonly accepted proxy for the publication quality: publication citation counts.
    To ensure a robust comparison we used the DOI of the published version for each $\star$rXiv publication, when available, to identify corresponding citation data.
    Openalex provides data on citation counts by year, as well as the total citation count as of the date database access.
    To compose the citation metrics for SI at horizon $\Delta = 36$ months we use two approaches.
    In the first one we consider the total citation counts as of access date to Openalex API, in the second one we aggregate the citations from the consecutive three years following the publication of the paper and used the total counts.
    To address temporal mismatches between publication and citation observation dates, we derive the correlation using a 90-day window.
    The adjustment accounted for differences in the scaling of SI and for variations in citation counts arising from the differing ages of papers (in each batch we compare citations between publications that have at most 90 days difference in posting dates).

    We observed that both approaches yield similar results (see Fig.~\ref{fig:scale_citations}, right and Fig. \ref{fig:scale_citations.appendix}, right).

    We consider two types of models, a linear regression model with L1-penalty and the Histogram-based Gradient Boosting Regression Tree (HBR) both implemented in \textit{scikit-learn}.
    For Lasso, we optimize over the penalty term value, for HBR we optimize over maximum tree depth and the number of iterations.
    We observe that HBR outperforms the Lasso model on the chosen measure of quality -- the average coefficient of determination of $J_\pi$ and $\delta_\pi$ models.

    For each pair of training / validation samples we first train models for all possible hyperparameters for $J_\pi$, then select 3 top performing models and for each of them train models for $\delta_\pi$.
    At the end we select the pair of models that deliver the best sum of $R^2$s for prediction of $J_\pi$ and $\delta_\pi$.

\begin{figure}[h]
        \centering
        \hspace*{-1cm}\includegraphics[width=1.2\textwidth]{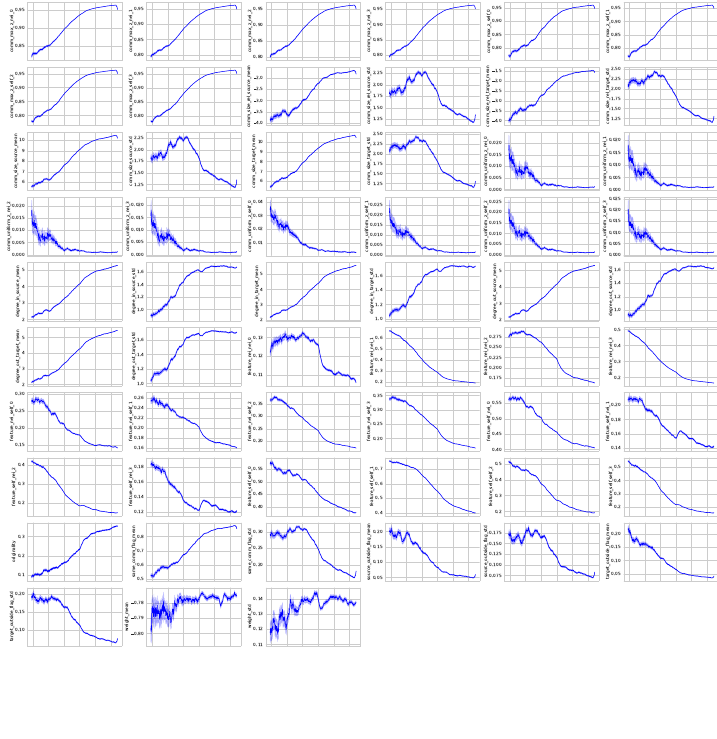}
        \caption{The mean of feature value as a function of time.}\label{fig:scale_features}
    \end{figure}

    As well as SI is non-stationary, the features (see Appendix~\ref{sec:appendix:b}, are also not stationary (see Fig. \ref{fig:scale_features}).

\begin{figure}[h]
        \centering
        \includegraphics[height=0.9\textheight]{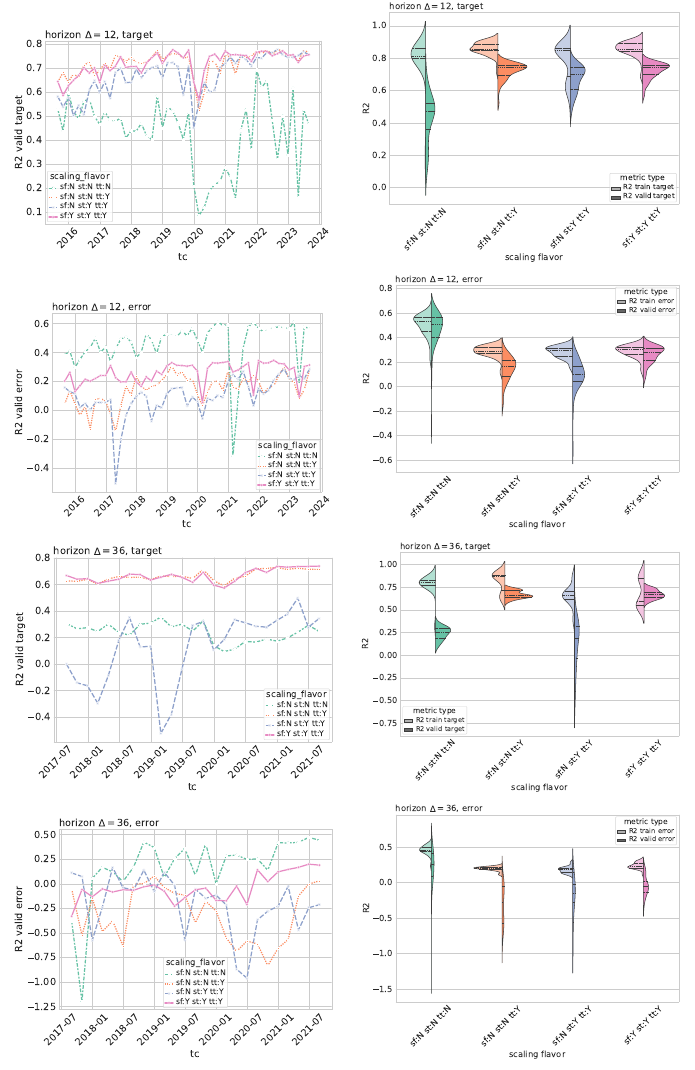}
        \caption{\textbf{Left}: line plots, representing coefficient of determination $R^2$ for the validation sample as a function of time. \textbf{Right}: violin plots of the distributions of coefficient of determination $R^2$ for the training and validation sample (left and right halves, lighter and darker hues of the same color correspondingly). Having in mind that \textit{sf} stands for scaling the features, \textit{st} - scaling the target, \textit{tt} - transforming the target, the four cases of scaling/transorfmation options are (NNN), (NNY), (NYY) and (YYY) in \textit{(sf, st, tt)} coordinates.}\label{fig:model.perf.sf}
    \end{figure}

    Hence, we first run a series of experiments with a reduced hyperparameter set ($\alpha \in \{10^{-4}\ldots 10^{-1}\}$ for Lasso, max depth=5, max iterations=100 for HBR) to decide whether to transform/scale the target variable and/or transform the features, we run the first optimization experiment for the four combinations of three possibilities (scale features, transform target, scale target): (n, n, n), (n, y, n), (n, y, y) and (y, y, y).
    The results are presented in Fig. \ref{fig:model.perf.sf}.

    We note that the SI predictions for target transform and transform-scale target and scale features have similar performances on the validation sample, while the rest of the combinations lag behind.
    The combination of transform-scale target and scale features delivers the best performance if we take into account the performance of risk predicting model as well.
    Hence, we include transformation $\ln (1 + \star)$ in the definition of $J_\pi$ and the features, as well as scaling them with respect to moving window average.
    We note that the performance of risk-predicting models is not stable.

\begin{figure}[h]
        \centering
        \includegraphics[width=0.9\textwidth]{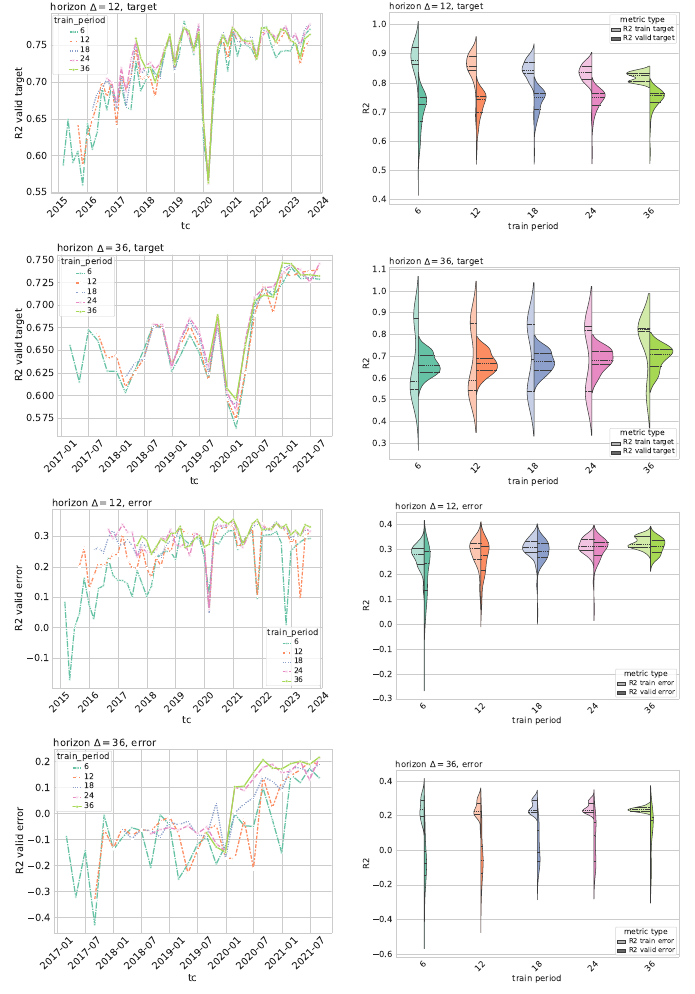}
        \caption{\textbf{Left}: line plots, representing coefficient of determination $R^2$ for the validation sample as a function of time. \textbf{Right}: violin plots of the distributions of coefficient of determination $R^2$ for the training and validation sample (left and right halves, lighter and darker hues of the same color correspondingly). Cases correspond to different lengths of training period in months. The target is transformed, both the target and the features are scaled.}\label{fig:model.perf.train.period}
    \end{figure}

    Next we run an experiment (see Fig.\ref{fig:model.perf.train.period}) on the same reduced hyperparameter set to decide on the effect on the length of the training period ($T_{train} \in {6, 12 , 18, 24, 36}$ [month]), which leads us to the conclusion that longer training periods have a positive effect on the target performance and a pronounced positive effect on the risk of impact across all values of prediction horizon.

    As the final step we select the best model type from the point of view of validation metrics over the hyperparameter set ($\alpha \in \{10^{-4}\ldots 10^{-1}\}$ for Lasso, max depth $\in  \{3, 4, 5, 6, 7\}$, max iterations $\in \{ 50, 100 \}$ for HBR).

    The out-of-sample predictive power of HBR models decreases with the increase of the prediction horizon from $0.76\pm 0.04$ for SI and $0.34\pm 0.02$ for SI prediction error for 6 month horizon to $0.69\pm 0.06$ for SI and $0.12\pm 0.09$ for 36 month horizon.



    \begin{figure}[h]
        \centering
        \includegraphics[height=0.9\textheight]{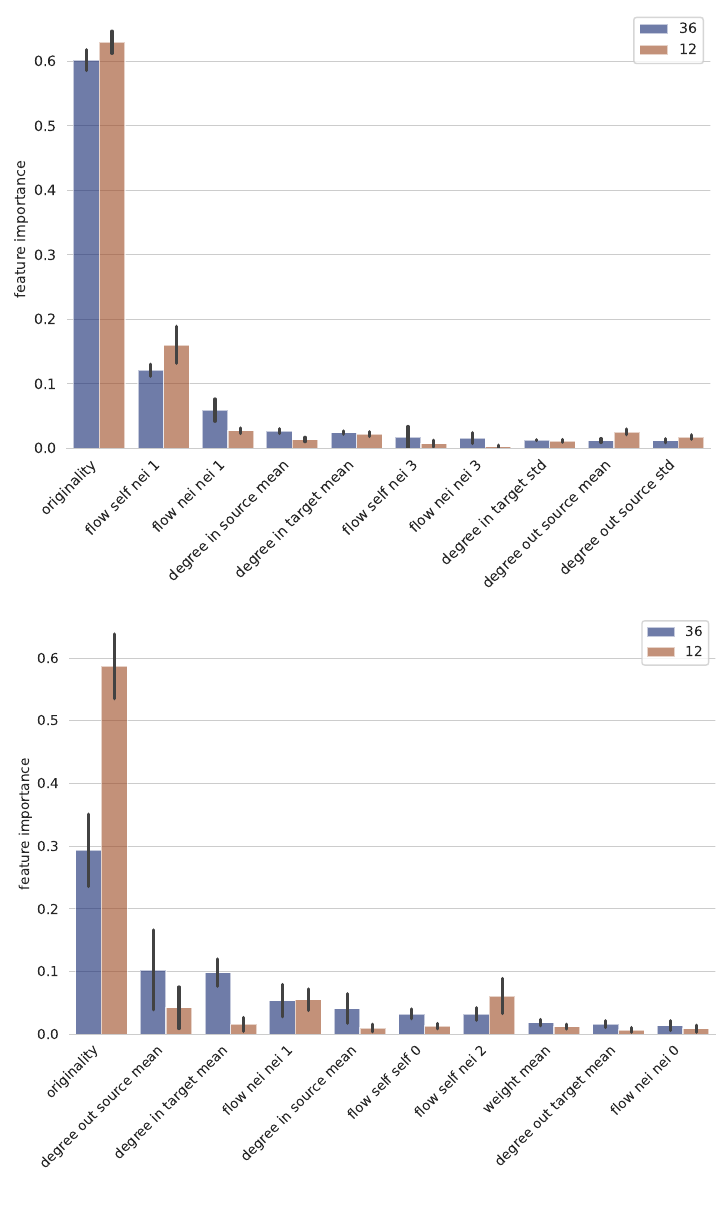}
        \caption{Average feature importances of optimals models of semantic impact $J_\pi$, top,  and its prediction error $\delta_\pi$, bottomm, for prediction horizons $\Delta = 12$ and 36 months.}\label{fig:model.imp.target.bar}
    \end{figure}

    \section{Features}\label{sec:appendix:b}
    The features used for SI and SI prediction error are derived for a subgraph $\gamma_\pi$ with respect to the complete graph $G_r(t)$ and fall into the following categories:
    \begin{itemize}
        \item \textbf{Originality}: inverse of the average weight of edges of $\gamma_\pi$ with respect to  $G_r(t)$ (see Eq. \ref{eq:originality}).
        \item \textbf{Degree-related metrics}: mean and standard deviation of in- and out- degrees computed separately for source and target vertices.
        \item \textbf{Diffusion-related metrics}: the fraction of measure remaining on $\gamma_\pi$ or its first-order neighbours after $k$ diffusion steps.
        Metrics are computed for up to 4 diffusion steps.
        \item \textbf{Community related measures}:
        \begin{itemize}
            \item the proximity of the distribution of vertex community attribution to a uniform distribution, expressed by p-value of Kolmogorov-Smirnov test of vertex community counts vs uniform distribution (A p-value $< 0.05$ indicates significant deviation from uniformity, whereas a p-value $\ge 0.05$ suggests the distribution may be uniform.).
            The metric is computed for subgraph $\gamma_\pi$ and its first-order neighbours.
            \item the mean and the standard deviation for the absolute and relative community sizes of $G_r(t)$ for the source and the target vertices of $\gamma_\pi$.
        \end{itemize}
        \item \textbf{Vertex novelty}: the mean and the standard deviation of the binary indicator denoting whether a vertex in $\gamma_\pi$ existed in $G_r(t)$ prior to the current time step, calculated for source and target vertex types.
        \item \textbf{Same community flag}: the mean and the standard deviation of the indicator variable specifying whether each source-target vertex pair shares a community.
    \end{itemize}

    The average feature importances are dominated by originality, while the second place is held by diffusion features (see Fig.\ref{fig:model.imp.target.bar}).

    \section{Covid effect}\label{sec:appendix:c}
    In this section, we demonstrate that COVID-19, as a strong external factor, disrupted the knowledge graph $G_r(t)$ and temporarily altered the distribution of SI.
    We focus on SI evaluated over a 36-month horizon.
    Similar results are expected for other horizon lengths.

\begin{figure}[h]
        \centering
        \includegraphics[width=0.9\textwidth]{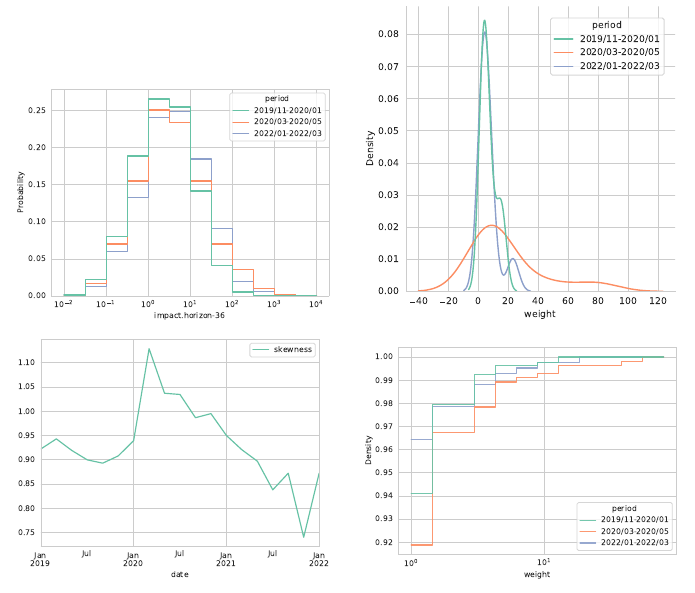}
        \caption{\textbf{Right}: top, SI distribution with prediction horizon of 36 months; bottom, the skewness of SI as a function of time. \textbf{Left}: top, KDE estimate of edge weight distribution, bottom, the cumulative distribution function of edge weight distribution. In all four figures, the samples are aggregated over two month periods.}\label{fig:covid}
    \end{figure}

    First, we observe that the shape of SI distribution for publications within a two-month interval changes in the periods immediately following January 2020.
    First, we observe that the shape of SI distribution for publications within a two-month interval changes in the periods immediately following January 2020.
    During the COVID-19 era, the distribution exhibits a fat tail for high impact values compared to previous or future intervals (see Fig.~\ref{fig:covid}, left, top).


    To further analyze this effect, we plot the skewness of the $J_\pi$ distribution as a function of time (see Fig.~\ref{fig:covid}, left, bottom).
    We observe a sharp increase in skewness, followed by a return to the baseline value.

    To determine whether a specific topic is responsible for perturbing the semantic graphs, we analyze the distribution of edge counts.
    We find that the edge count distribution widens, develops a fat tail, and subsequently returns to its original form.
    This pattern is also evident in the cumulative density function plot below, (top and bottom of right of Fig.~\ref{fig:covid}).

    Finally, we examine the most frequent edges across five two-month intervals spanning the early COVID-19 period.
    We find that COVID-related edges dominate the distribution, confirming the disruption (see Table~\ref{table:covid}).

    \begin{table}[!h]
            \caption{The statistics of most popular knowledge graph edges among publications aggregated by two-month periods.}

            \begin{tabular}{c|c|c|c|c|c}
                \toprule
                period & subject.id        & subject         & object.id       & object       & count \\
                \midrule
                \makecell{2019/11- \\\quad2020/01}       & NCBITaxon:9606      & Homo sapiens       & mesh:D009369  & Neoplasms & 16    \\
                & mesh:D009369 & Neoplasms        & NCBITaxon:9606  & Homo sapiens & 15    \\
                & wikidataId:Q42918 & mutation & NCBITaxon:9606  & Homo sapiens & 10     \\
                & wikidataId:Q26972  & gene expression          & NCBITaxon:9606  & Homo sapiens & 6     \\
                & wikidataId:Q7020 & genome & NCBITaxon:9606 & Homo sapiens & 5 \\

                \midrule

                \makecell{2020/01-       \\\quad2020/03}       & NCBITaxon:9606    & Homo sapiens    & mesh:D009369 & Neoplasms     & 19    \\
                & mesh:D000086382 & COVID-19 & NCBITaxon:9606 & Homo sapiens & 16 \\
                & NCBITaxon:9606 & Homo sapiens & mesh:D000086382 & COVID-19 & 14 \\
                & mesh:D000086382 & COVID-19 & mesh:D045169 & \makecell{Severe Acute \\ Respiratory Syndrome}       & 11   \\
                & mesh:D000086382 & COVID-19 & wikidataId:Q148 & \makecell{People's Republic \\of China} & 10 \\

                \midrule

                \makecell{2020/03- \\\quad2020/05} & NCBITaxon:9606 & Homo sapiens & mesh:D000086382 & COVID-19 & 78 \\
                & mesh:D000086382   & COVID-19        & NCBITaxon:9606  & Homo sapiens & 44    \\
                & NCBITaxon:9606 & Homo sapiens & mesh:D045169 & \makecell{Severe Acute \\ Respiratory Syndrome} & 18 \\
                & mesh:D000086382 & COVID-19 & mesh:D045169 & \makecell{Severe Acute \\ Respiratory Syndrome} & 13 \\
                & mesh:D045169   & \makecell{Severe Acute        \\ Respiratory Syndrome} & NCBITaxon:9606 & Homo sapiens & 10 \\

                \midrule

                \makecell{2020/09-       \\\quad2020/11} & NCBITaxon:9606 & Homo sapiens & mesh:D000086382 & COVID-19 & 82 \\
                & mesh:D000086382 & COVID-19 & NCBITaxon:9606 & Homo sapiens & 31 \\
                & NCBITaxon:9606 & Homo sapiens & mesh:D045169 & \makecell{Severe Acute \\ Respiratory Syndrome} & 20 \\
                & mesh:D000086382 & COVID-19 & mesh:D045169 & \makecell{Severe Acute \\ Respiratory Syndrome} & 14 \\
                & mesh:D045169 & \makecell{Severe Acute \\ Respiratory Syndrome} & mesh:D000086382 & COVID-19 & 8 \\

                \midrule

                \makecell{2021/07- \\\quad2021/09} & NCBITaxon:9606 & Homo sapiens & mesh:D000086382 & COVID-19 & 64 \\
                & mesh:D000086382 & COVID-19 & NCBITaxon:9606 & Homo sapiens & 29 \\
                & NCBITaxon:9606 & Homo sapiens & mesh:D045169 & \makecell{Severe Acute \\ Respiratory Syndrome} & 20 \\
                & NCBITaxon:9606 & Homo sapiens & mesh:D009369 & Neoplasms & 8 \\
                & mesh:D045169 & \makecell{Severe Acute \\ Respiratory Syndrome} & NCBITaxon:9606 & Homo sapiens & 8 \\
                \bottomrule
            \end{tabular}
            \label{table:covid}
    \end{table}

\end{appendices}

\end{document}